\documentclass{elsart3p}

\usepackage{graphicx}

\usepackage{amssymb}

\begin{document}

\begin{frontmatter}



\title{PEBS - Positron Electron Balloon Spectrometer}

\author{P. von Doetinchem, H. Gast, T. Kirn, G. Roper Yearwood, S. Schael} 
\address{{I. Physikalisches Institut B, RWTH Aachen, Germany}}


\begin{abstract}
The best measurement of the cosmic ray positron flux available today was performed by the
HEAT balloon experiment more than 10 years ago. 
Given the limitations in weight and power consumption for balloon experiments, a novel approach was needed
to design a detector which could increase the existing data by more than a factor of 100.\\
Using silicon photomultipliers for the readout of a scintillating fiber tracker and of an imaging electromagnetic calorimeter, 
the PEBS detector features a large geometrical
acceptance of $2500~cm^2 sr$ for positrons, 
a total weight of $1500\,kg$ and a power consumption of $600\,W$. The experiment is 
intended to measure cosmic ray particle spectra for a period of up to 20 days at an altitude of $40\,km$ circulating the North or South Pole.\\
A full Geant 4 simulation of the detector concept has been developed and key elements have
been verified in a testbeam in October 2006 at CERN. 
\end{abstract}

\end{frontmatter}

\section{Introduction}
\label{sec::intro}
Among the most intriguing open questions in modern physics is the
nature of the dark matter, that has been shown to contribute around
22~\% to the total energy density of the universe. Certain models,
such as supersymmetric extensions to the standard model of particle
physics, predict a new particle, the neutralino, which has all the
properties required of a dark matter candidate. It will form halos
around galaxies and annihilate pairwise into known
particles. At the end of their decay chains, positrons and electrons
will be produced in equal numbers. Since there is no other known primary source of
positrons in the Galaxy, positrons provide an excellent probe for
the indirect detection of dark matter. The observation of an excess
over the expected secondary flux by the HEAT\cite{cite::heat} and
AMS-01\cite{cite::ams01} experiments has sparked some excitement but it needs
to be confirmed by a precise measurement.

\section{Detector description}
\label{sec::overview}
An experiment designed to measure the positron component in the cosmic
rays has to fulfill several crucial requirements:
\begin{itemize}
\item The geometrical acceptance needs to be larger than $1000\,cm^2\,sr$ due
to the small flux of positrons.
\item A suppression of the predominant proton background
of $10^6$ has to be achieved.
\item A good momentum resolution is necessary for charge sign
determination and subsequent electron suppression.
\end{itemize}
The PEBS detector has been designed to meet these requirements. We have
conducted a full simulation of the behavior of the experiment using
the Geant4 package\cite{cite::g4}. In addition, key elements have
been verified in a testbeam in October 2006 at CERN. 

\begin{figure}
\begin{center}
\includegraphics[width=5cm,angle=90]{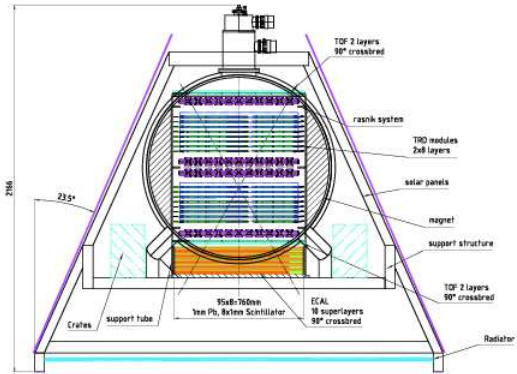}
\end{center}
\caption{Cutaway mechanical drawing of the PEBS design including
support structure and solar panels.}
\label{fig::pebs_mechanical}
\end{figure}

A mechanical drawing of the PEBS detector including
support structure, electronics crates and solar panels can be seen in
figure~\ref{fig::pebs_mechanical}. The apparatus has an overall height
of $2.17\,m$, a length of $3.23\,m$ and a width of $2.43\,m$.

A magnetic field of mean flux density $B=1\,T$ is created by two
superconducting Helmholtz coils, located in a helium cryostat.
The curvature of a charged particle's trajectory in this field is measured by
a scintillating fiber tracker with silicon photomultiplier readout. A
transition radiation detector(TRD), located between the tracker
super-layers, and an electromagnetic calorimeter at the bottom of the
experiment provide rejection power against protons. The performance of
these components is evaluated in this section.

Scintillator panels above and below the tracker act as a
time-of-flight system (TOF) and are used for triggering purposes.

\subsection{Mission parameters}
\label{sec::mission}
Earth's atmosphere prohibits a measurement of $GeV$-range cosmic rays
on the ground. As an interesting alternative to space-based measurements, a
high-altitude balloon is chosen. Mission durations of around 40~days can be reached by
traveling with the circular arctic winds around the North or South
Pole\cite{cite::cream}.

The geometrical acceptance of the detector is limited by the weight
and power constraints imposed by the carrier system. The most
important contributions to the overall weight are the
magnet weight and the weight of the calorimeter with $850\,kg$ and
$550\,kg$ respectively. The power consumption is dominated by the
$260\,W$ needed for the tracker which has roughly 50000 individual
readout channels.

\begin{figure}
\begin{center}
\includegraphics[width=8cm,angle=0]{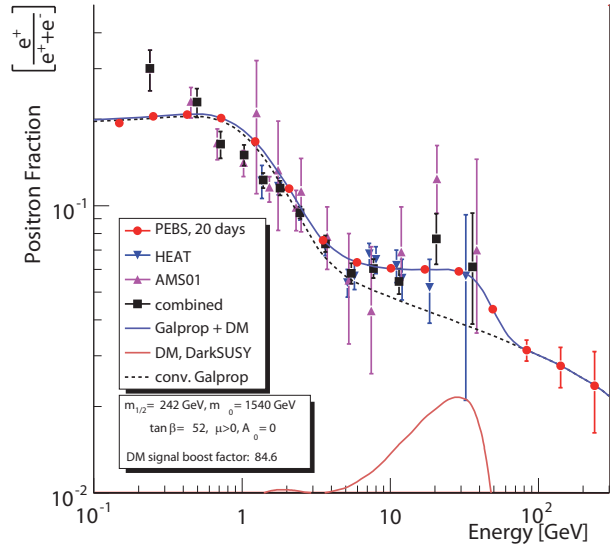}
\end{center}
\caption{Existing data for the cosmic-ray positron fraction from
HEAT\cite{cite::heat} and AMS01\cite{cite::ams01} together with the 
projected PEBS data. Model predictions have been estimated using
DarkSusy\cite{cite::darksusy}. The signal process shown is based on an MSSM model with
neutralino dark matter.}
\label{fig::pebs_performance}
\end{figure}

The magnet dimensions of $80\times 80\times 80\,cm^3$ allow
for a maximum acceptance of $4020\,cm^2sr$ \cite{cite::sullivan}. For
an overall detector length of $1\,m$ and an effective tracker width of
$76\,cm$, an acceptance of $2460\,cm^2sr$ is
calculated. Figure~\ref{fig::pebs_performance} shows the statistical errors
on the positron fraction achievable with the PEBS acceptance in a
20-day campaign, as compared to the currently available data.

\subsection{Tracker}
\label{sec::tracker}
The tracking device will consist of scintillating fibers grouped into
modules and read out by silicon photomultipliers (SiPMs). A module
comprises two stacks of round fibers of $250\,\mu{}m$ diameter,
128~fibers wide and four fibers high, in the tightest arrangement. The
stacks are held apart by two carbon fiber skins with Rohacell foam in
between. Using
scintillating fibers, the material budget in the particles' flight
path through the tracker does not exceed
$6\,\%$ of a radiation length, while the TRD will contribute another
$6\,\%$. The modules will be grouped into eight layers, two of those
being located at the entrance and exit of the tracking device
respectively, and four in the center. In this arrangement, the
uncertainty in momentum determination is minimized\cite{cite::gluckstern}.

\begin{figure}
\begin{center}
\includegraphics[width=8cm,angle=0]{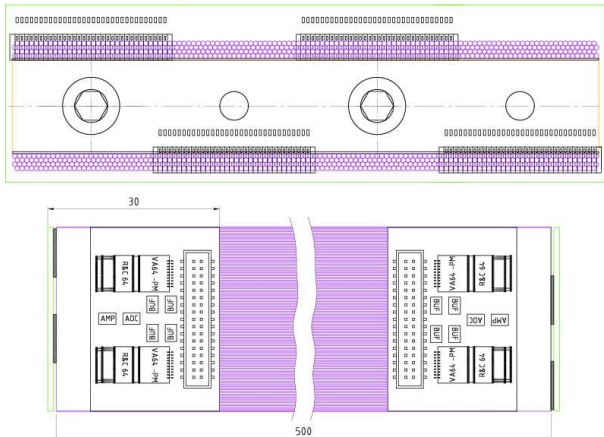}
\end{center}
\caption{({\it top}) Front view of a tracker module showing the fibers
with corresponding SiPM arrays, mounted to a support structure
composed of carbon fiber skins held apart by Rohacell foam. ({\it
bottom}) Top view of a tracker module including a design for the front
end hybrids to read out the SiPMs.}
\label{fig::tracker_module}
\end{figure}

Silicon photomultipliers\cite{cite::sipm} have the virtues of being insensitive to
magnetic fields, having high quantum efficiency, as well as
compactness and auto-calibration. They will therefore be used to
detect the photons trapped in the scintillating fibers and will be
read out by a dedicated VA chip. Figure~\ref{fig::tracker_module}
illustrates the readout scheme. Arrays containing 32~silicon
photomultiplier columns each are located at alternating ends of the fiber
bundles. The remaining end of each fiber is covered by a reflective foil to
increase the light yield by a factor of roughly $1.6$. Four
fibers in one column are then optically connected to one SiPM column.
The weighted cluster mean from amplitudes in adjacent SiPMs columns will be
calculated to pinpoint the intersection of a trajectory with a fiber
module.

A prototype of the tracking device, built of two fiber bunches, each
consisting of ten stacks of three square fibers of $300\,\mu{}m$
width, has been subjected to a $10\,GeV$ proton testbeam at the
CERN~T9 beamline. The detailed analysis of the data gathered is
presented elsewhere\cite{cite::gregorio}. 

\begin{figure}
\begin{center}
\includegraphics[width=8cm,angle=0]{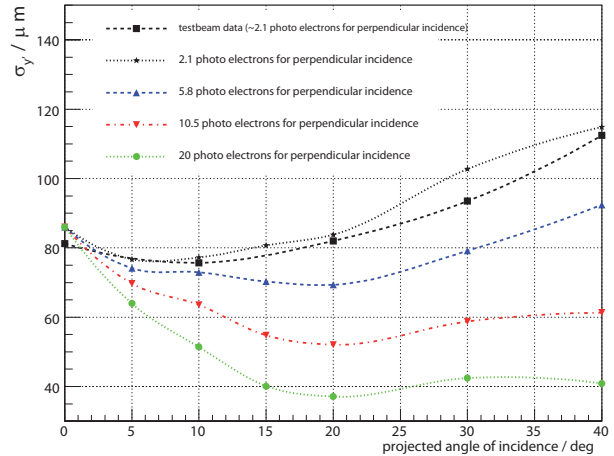}
\end{center}
\caption{Spatial resolution for a bundle of fibers of $300\,\mu{}m$
width from testbeam data and Monte Carlo simulations. Testbeam data
obtained with a fiber bundle without reflective foil and Photonique
SSPM-050701GR SiPM
readout are plotted using square markers. Results from
Monte Carlo simulations are added to study the behavior for improved
photo electron yields. A yield of $5.8$ photo electrons was reached in
the testbeam with SSPM-0606EXP SiPMs and reflective foil, but only
data at $0^\circ$ were taken in this configuration.}
\label{fig::spatialresolution}
\end{figure}

A dedicated Monte Carlo simulation, again using the Geant4 package,
has been developed for comparison to and generalization of the
testbeam results. A key question to be answered was the spatial
resolution obtained with a fiber module as a function of the mean photo
electron yield $n_{p.e.}$ of the fiber-SiPM chain. 
Figure~\ref{fig::spatialresolution}
shows the result. The spatial resolution $\sigma_{y^\prime}$ is plotted
for different values of $n_{p.e.}$ and depending on the angle $\alpha$ of incidence
of a particle, projected into the bending plane of the
magnet. $\sigma_{y^\prime}$ is the resolution along the axis
perpendicular to the fibers. Since the beam telescope used in the
testbeam measured the coordinate $y$ perpendicular to the direction of
incidence $z$, $\sigma_{y^\prime}$ is calculated from the measured
$\sigma_y$ and the positioning accuracy $\sigma_z=10\,\mu{}m$ as
follows:
\begin{equation}
\sigma_{y^\prime}=\sigma_y \cos\alpha\oplus
\sigma_z\sin\alpha
\end{equation}
For the photo electron yield achieved in the testbeam, a spatial
resolution of $72\,\mu{}m$ is obtained at the mean projected angle of
incidence, which is $\bar{\alpha}=11^\circ$ for the PEBS geometry.

The full PEBS detector simulation was then used to determine the momentum
resolution, achievable with the tracker design for a photo electron
yield corresponding to the one reached in the testbeam. Muons in the momentum
range up to $100\,GeV$ were simulated and the reconstructed momentum
resolution was parameterized as
\begin{equation}
\sigma\left(\frac{p_{\mathrm{MC}}}{p_{\mathrm{rec}}}\right) = 
a_{\mathrm{msc}}\oplus b_{\mathrm{res}}\cdot p_{\mathrm{MC}}
\end{equation}
where $p_{\mathrm{MC}}$ and $p_{\mathrm{rec}}$ denote generated and
reconstructed momentum respectively.
In the current configuration, the
simulation yields values of $a_{\mathrm{msc}}=2\,\%$ and
$b_{\mathrm{res}}=0.188\,\%\,/\,GeV$.

\subsection{Electromagnetic calorimeter}
\label{sec::ecal}
\begin{figure}
\begin{center}
\includegraphics[width=8cm,angle=0]{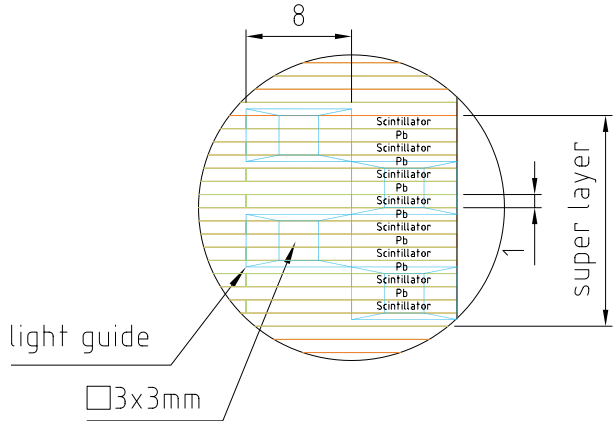}
\end{center}
\caption{Mechanical drawing showing a part of an ECAL
super-layer. SiPMs will be mounted to the end of the lightguides.}
\label{fig::ecaldrawing}
\end{figure}

A sandwich calorimeter for three-dimensional shower reconstruction has been designed
to provide rejection power against the predominant proton component in
the cosmic rays. It comprises 80~layers consisting of $1\,mm$ lead 
interleaved with layers of $8 \times 1\,mm^2$  scintillating bars.  They are read out 
by $3\times{}3\,mm^2$ SiPMs with 8100 pixels which
are connected to the fibers using light-guides as sketched in
figure~\ref{fig::ecaldrawing}. Ten layers are grouped
into a super-layer and super-layers are placed with alternating
direction.
The total depth of the calorimeter is $14.3$~radiation lengths.

\begin{figure}
\begin{center}
\includegraphics[width=8cm,angle=0]{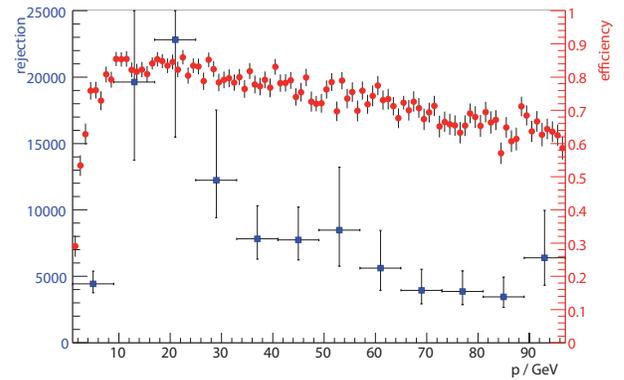}
\end{center}
\caption{Proton rejection and electron efficiency of the ECAL alone,
as obtained from the full detector simulation. Squares denote proton
rejection (left axis) while circles represent the corresponding
electron efficiency (right axis).}
\label{fig::ecal}
\end{figure}

A preliminary cut-based analysis has been performed to study the proton rejection of
this setup. For each event, a shower fit using a standard Gamma
function parameterization has been performed and the following
variables have been used to distinguish 
positrons from protons:
\begin{itemize}
\item $E/p$-matching: The fitted shower energy has to match the
reconstructed momentum in the tracker.
\item $t_\mathrm{max}$: The fitted shower maximum must be at an
appropriate depth in the calorimeter.
\item The ratio of shower energy within one Moli\`ere radius from the
shower axis has to be
characteristic for an electromagnetic shower.
\item The angle between the reconstructed track and shower axis must
be small.
\end{itemize}
The resulting proton rejection and positron efficiencies are depicted
in figure~\ref{fig::ecal}.

Proton rejections of the order of 5000 can
easily be achieved already with this rather
coarse method. The corresponding electron efficiency is around $70\,\%$.

\subsection{Transition radiation detector}
\label{sec::trd}
The design of the transition radiation detector is based on the one
constructed for the AMS-02 experiment on the International Space
Station\cite{cite::ams02trd}. The TR x-ray photons are generated in a
$2\,cm$ thick irregular fleece radiator made of polyethylene and
polypropylene. They are subsequently detected in proportional wire
chambers in the form of straw tubes made of
aluminized kapton foils which have an inner diameter of $6\,mm$ and
are filled with an $80:20$ mixture of Xe/CO$_2$. The straw tubes are
grouped into modules and eight layers each are placed in the gaps
above and below the central tracking layers.

\begin{figure}
\begin{center}
\includegraphics[width=8cm,angle=0]{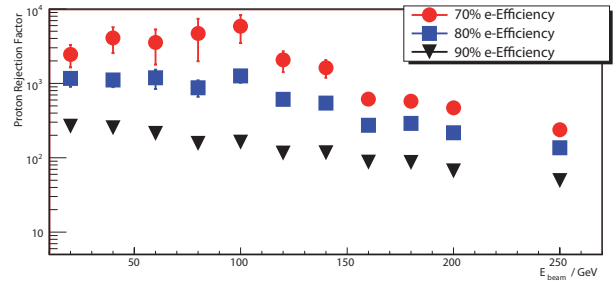}
\end{center}
\caption{Proton rejection of the TRD alone, for various values of the
electron efficiency.}
\label{fig::trd}
\end{figure}

Detailed performance studies using both Monte Carlo and testbeam data
have been conducted. The proton rejection yielded by the TRD is
depicted in figure~\ref{fig::trd}. It reaches a value of 1000 at
$80\,\%$ electron efficiency in the interesting energy range.

More information on the design and performance of the TRD can be found elsewhere\cite{cite::thomas}.

\section{Conclusion}
\label{sec::conclusion}
We have presented a design study to construct a balloon-borne
cosmic ray spectrometer to measure the positron
fraction. Scintillating fibers with SiPM readout are used as key
components for the tracker. 
This large high-resolution tracker inside a magnetic
field of $1\,T$ will allow a precise measurement of the positron spectrum, thus providing valuable
information for the indirect search for dark matter. The necessary
proton rejection of the order of $10^6$ can be achieved by the combination of
a 3D-imaging calorimeter and a 16-layer transition radiation detector. 



\end{document}